\documentstyle[preprint,pra,aps,epsf]{revtex}
\tightenlines

\begin{document}
\draft
\preprint{CfA No. 4767}
\title{Variational calculations on the hydrogen molecular ion}
\author{J. M. Taylor, Zong-Chao Yan, A. Dalgarno, and J. F. Babb}
\address{
Institute for Theoretical Atomic and Molecular Physics,\\
Harvard-Smithsonian Center for Astrophysics,\\
60 Garden Street, Cambridge, MA 02138
}

\maketitle
%
\begin{abstract}
We present high-precision non-relativistic variational calculations of
bound vibrational-rotational state energies for the $\mbox{H}_2{}^+$ and
$\mbox{D}_2{}^+$ molecular ions in each of the lowest electronic states of
$\Sigma_g$, $\Sigma_u$, and $\Pi_u$ symmetry.  The calculations are
carried out including coupling between $\Sigma$ and $\Pi$ states but
without using the Born-Oppenheimer or any adiabatic approximation.
Convergence studies are presented which indicate that the resulting
energies for low-lying levels are accurate to about $10^{-13}$.  Our
procedure accounts naturally for the lambda-doubling of the $\Pi_u$
state.
\end{abstract}
\pacs{PACS numbers: 31.15.Ar, 31.15.Pf, 33.15.Fm}


\narrowtext
\section{INTRODUCTION}

There are many calculations of bound state energies of the hydrogen
molecular ion $\mbox{H}_2{}^+$ using the Born-Oppenheimer
approximation or various adiabatic approximations and there are a
number of studies that investigate deviations of energies from the
Born-Oppenheimer values.  The present work is a systematic high
precision nonadiabatic\footnote{We would prefer to use the term
`batic, which we coined to avoid the double negative implied in
nonadiabatic, but clarity must yield to convention.} study of
$\mbox{H}_2{}^+$ and $\mbox{D}_2{}^+$ in each of the lowest electronic
states of $\Sigma_g$, $\Sigma_u$, and $\Pi_u$ symmetry carried out
using variational basis sets. It is motivated by recent precise
experimental spectroscopy of Rydberg states of the hydrogen and
deuterium molecules that has led to accurate experimental values of
the the electric dipole polarizability of the corresponding molecular
ions in their ground states~\cite{JacFisFeh97}.  These experiments
were followed by several papers detailing various nonadiabatic
calculations of the electric dipole
polarizability~\cite{SheGre98,BhaDra98,Mos98,Cla98}.  The present
paper is the first in a series.  We are using the eigenstates studied
in the present work in a study of the electric dipole sum rules for
$\mbox{H}_2{}^+$ and $\mbox{D}_2{}^+$, including the polarizability.

Several  investigators have  performed nonadiabatic calculations on the
ground electronic state of $\mbox{H}_2{}^+$ since Hunter and
Pritchard~\cite{HunPri67a} and Ko\l{}os~\cite{Kol69} reported the
first precision calculations. The most accurate calculations used
variational basis set methods~\cite{Bis89,Mos90},
variation-perturbation methods~\cite{WolPol86,WolOrl91}, and
artificial channel scattering methods~\cite{Mos93a,Mos93b}.
Variational basis set calculations can be in principle quite  accurate
but appear to have been applied only to the lowest-lying eigenvalues
of the $\Sigma_g$ symmetry. The variation-perturbation and the
artificial channel methods yield energies for all of the
vibration-rotational levels and have been applied to the states of
$\Sigma_g$ and $\Sigma_u$ symmetry.  There are other approaches 
applied to the $\Sigma_g$ symmetry that
have not yet reported precision as great as those mentioned above such
as the adaptive finite element method~\cite{AckShe96}, the generator
coordinate method~\cite{RibTolPiz83}, quantum Monte
Carlo~\cite{BreMelMor97} and perturbative approaches~\cite{BabDal91}.
Energy calculations up to 1980 were reviewed by Bishop and
Cheung~\cite{BisChe80} and a useful, more general review covering up
to 1995 can be found in~\cite{LeaMos95}.


\section{THEORY}
In this section we derive the Hamiltonian and introduce the basis sets
we used.  Other derivations can be found in
Refs.~\cite{JepHir60,KolWol63,HunGraPri66,CarKen84,WolPol86,MosSad89}.
Some of the operators we use were introduced in those references and
Ref.~\cite{Joh41}.  Our intention is to avoid writing explicit matrix
elements until the last steps and the spirit of the present derivation
is closest to the derivations in
Refs.~\cite{JepHir60,HunGraPri66,PacHir68}.
\subsection{Hamiltonian}

In a space-fixed frame and 
with the center of mass motion removed the Hamiltonian for the
homonuclear one-electron diatomic molecule is
\begin{equation}
\label{ham}
H = -\case{1}{2}M^{-1}\nabla_R^2 
 -[\case{1}{2}+\case{1}{8}M^{-1}]\nabla^2  + V({\bf r},{\bf R}) ,
\end{equation}
where
\begin{equation}
\label{potential}
V ({\bf r},{\bf R} ) = 
  -\frac{1}{| {\bf r} - {\case 1 2}{\bf R} |}
  -\frac{1}{| {\bf r} + {\case 1 2}{\bf R} |} 
 + \frac{1}{R} 
\end{equation}
and $M=\case{1}{2}M_n$, with $M_n$ the nuclear mass, ${\bf r}$ the
position vector of the electron from the midpoint of the vector $\bf
R$ joining the nuclei, and $R=|{\bf R}|$.  We use atomic units
throughout.  
The electronic (cartesian) coordinates are to be held fixed in the space-fixed
frame in carrying out the derivatives  in the gradient operator
$\nabla_R$ appearing  in  Eq.~(\ref{ham})~\cite{VanVle29,Bun68}.

Following Ref.~\cite{LefFie86} we introduce the rotational angular
momentum ${\cal R}$ implicitly expressing the Hamiltonian in a
rotating molecular fixed frame. The nuclear kinetic energy is written
as
\begin{equation}
\label{nke-pre}
-\frac{\nabla_R^2}{2 M}  = \frac{1}{2M R^2}
  \left( -\frac{\partial}{\partial R} R^2\frac{\partial}{\partial R}
 + {\cal R}^2 \right) .
\end{equation}
Defining a rotational Hamiltonian
\begin{equation}
\label{hrot}
H_{\rm rot} =  \frac{{\cal R}^2}{2MR^2}
\end{equation}
we write
\begin{equation}
-\frac{\nabla_R^2}{2 M} =
 -\frac{1}{2M R^2}
  \frac{\partial}{\partial R} R^2\frac{\partial}{\partial R}
  + H_{\rm rot} ,
\end{equation}
where the three spherical polar coordinates comprised of $R$ and the
two angles (contained in the ${\cal R}^2$ operator of $H_{\rm rot}$)
contain the information on the orientation of the molecular fixed frame
with respect to the space fixed frame.

Since here we are ignoring electron and nuclear spins, the total
angular momentum is ${\bf N}={\bf {\cal R}}+{\bf L}$, where $\bf L$ is
the electronic angular momentum.  Using ${\bf {\cal R}}={\bf N}-{\bf
L}$, we replace ${\bf \cal R}^2$ in Eq.~(\ref{hrot}) giving
\begin{equation}
\label{hrot-expand}
H_{\rm rot} =  \frac{1}{2MR^2}
  {({\bf N}-{\bf L})^2} = \frac{1}{2MR^2} (N^2 + L^2
  - N^- L^+  - N^+ L^-  - 2 N_z L_z) ,
\end{equation}
where the superscripts on $L^+$ and $L^-$ and subscript $z$ on $L_z$
refer to the components in the molecule-fixed frame~\cite{LefFie86}.

Changing  the electron coordinates from cartesian to
prolate spheroidal coordinates ($\lambda,\mu,\chi$),
we have
$r=|{\bf r}|=\case{R}{2}(\lambda^2+\mu^2-1)^{1/2}$.
The operator $\frac{\partial}{\partial R}$
in (\ref{nke-pre}) is taken with 
the electronic (prolate spheroidal)
coordinates held fixed in 
the molecular fixed frame
and can be expressed as
\begin{equation}
\label{partial}
\frac{\partial}{\partial R} =
\frac{\partial}{\partial R}\left.\right)_{\lambda,\mu}
 - \frac{\partial r}{\partial R}\frac{\partial}{\partial r} ,
\end{equation}
where the term $ \frac{\partial}{\partial R}$ on the LHS
of Eq.~(\ref{partial}) refers to the derivative
with the electronic (cartesian) coordinates held fixed
as in Eq.~(\ref{nke-pre}).

Using the RHS of Eq.~(\ref{partial}) in Eq.~(\ref{nke-pre}) we can 
write the kinetic energy operator as
\begin{equation}
\label{nke}
-\frac{\nabla_R^2}{2M} 
   =  \frac{1}{2M} \left[-\frac{\partial^2}{\partial R^2}
     - \frac{2}{R}\frac{\partial}{\partial R}  
   + \frac{2Y}{R^2} \frac{\partial}{\partial R}R
     - \frac{r^2}{R^2} p_r^2 
   + H_{\rm rot} \right],
\end{equation}
where 
\begin{equation}
p_r^2 =   -\frac{1}{r^2}\frac{\partial}{\partial r} r^2
           \frac{\partial}{\partial r} 
\end{equation}
and 
\begin{equation}
\label{Y-spheroidal}
Y= r \frac{\partial}{\partial r} 
\end{equation}
and it is now understood that 
the electronic (prolate spheroidal) coordinates are held fixed
where appropriate.

We use the expression
\begin{equation}
-p_r^2 - \frac{L^2}{r^2}
 = \nabla^2
\end{equation}
to combine Eq.~(\ref{nke}) and (\ref{hrot-expand}), yielding
\begin{equation}
\label{KE-final}
-\frac{\nabla_R^2}{2M} 
   =  \frac{1}{2M} \left[-\frac{\partial^2}{\partial R^2}
     - \frac{2}{R}\frac{\partial}{\partial R}  
   + \frac{2Y}{R^2} \frac{\partial}{\partial R}R
     + \frac{r^2}{R^2} \nabla^2 
   + \frac{1}{R^2}  (N^2 
  - N^- L^+  - N^+ L^- - 2 N_z L_z) \right] .
\end{equation}

We define for later use the coupling term 
\begin{equation}
\label{sigpi}
\frac{1}{2 M R^2} (-N^- L^+  - N^+ L^- ) 
\end{equation}
that enters
from Eq.~(\ref{KE-final})
into the Hamiltonian.

The potential
energy is given in terms of the  prolate 
spheroidal coordinate system $(\lambda,\mu,\chi)$ by 
\begin{equation}
V (\lambda,\mu,R) =\frac{1}{R}   -\frac{4\lambda}{R(\lambda^2-\mu^2)},
\end{equation}
and the electronic kinetic energy operator  by 
\begin{equation}
\nabla^2  = 
 (4/R^2)[X + 
 (\lambda^2-1)^{-1} (1-\mu^2)^{-1}\partial^2/\partial \chi^2] ,
\end{equation}
where
\begin{equation}
X=( \lambda^2 - \mu^2)^{-1}  
  [(\partial /\partial\lambda )(\lambda^2-1)\partial /\partial\lambda
  + 
  (\partial /\partial\mu )(1-\mu^2)\partial /\partial\mu] .
\end{equation}
and the operator $Y$, Eq.~(\ref{Y-spheroidal}), becomes
\begin{equation}
Y=( \lambda^2 - \mu^2)^{-1}  
  [\lambda(\lambda^2-1)\partial /\partial\lambda
  + 
  \mu(1-\mu^2)\partial /\partial\mu] .
\end{equation}
The terms in $\bf L$ can be reexpressed in the 
$(\lambda,\mu,\chi)$
coordinates, see for example Ref.~\cite{DalMcC57}.

The remainder of the Hamiltonian derivation  follows that
of, for example~\cite{WolPol86},  and in this way the
Hamiltonian reduces to effective matrix elements that
may be evaluated as integrals over $\lambda$, $\mu$, and $\chi$.


\subsection{Basis sets and trial functions}
For the electronic states of $\Sigma_g$, $\Sigma_u$
and $\Pi_u$ symmetry investigated here we used a basis set composed
of functions of the form~\cite{MosSad89}
\begin{equation}
\label{elec-basis}
\Phi_{bc}^{\Lambda p} (\lambda,\mu,\chi) = (\lambda^2-1)^{|\Lambda|/2} 
                      L_b^{|\Lambda|} [\alpha (\lambda-1)]
               \exp [-\case{1}{2}\alpha (\lambda-1)]
            P_c^{|\Lambda|} (\mu) \exp (i\Lambda\chi) ,
\end{equation}
with $b=0,...,B$ and $\alpha$ a nonlinear parameter.  We used values
of $\Lambda=-1,0$, and 1. The values $|\Lambda|=0$ and 1 correspond,
respectively, to $\Sigma$ and $\Pi$ states.  For the $\Sigma_g$ symmetry
$c=0,2,..,2C$ and $p=g$, for the $\Sigma_u$ and $\Pi_u$
symmetries $c=1,3,...,2C+1$ and $p=u$, and for the $\Pi_g$ symmetry
$c=2,4,...,2C+2$ with $p=g$.

The trial function for a particular set of states specified
by $\Lambda$, $p$, and $N$ has the form
\begin{equation}
\label{tf}
\Psi_{\Lambda p N} (\lambda,\mu,\chi,R) 
      = \sum_{s[bcd]=1}^S k_{s[bcd]}  
       \Phi_{bc}^{\Lambda p}(\lambda,\mu,\chi)  \chi_d (R)
\end{equation}
where $\Phi_{bc}^{\Lambda p}$ is given in Eq.~(\ref{elec-basis})
and where $S= (B+1)(C+1)(D+1)$. The index $s\equiv [bcd]$ was filled
in the order $[\{b,\{c,\{d\}\}\}]$, where $\{b\}$, for example,
indicates a loop over all possible values of the index $b=0,...,B$.
The vibrational basis functions were of the form 
\begin{equation}
\label{vib-basis}
\chi_d (R) = (1/R) (\gamma R)^{(\beta+1)/2} L_d^\beta (\gamma R)
   \exp (-\case{1}{2}\gamma R) ,
\end{equation}
with $d=0,...,D$.  
The vibrational state quantum numbers were
identified with levels in the spectrum resulting from the
diagonalization. 
The eigenvalues approach the exact eigenenergies behaving as
expected by the Hylleraas-Undheim theorem~\cite{New66}.

Laguerre polynomials were used in the electronic basis 
because  the integrals involved could be solved in
closed form. Other possibilities explored such as Hermite polynomials 
did not offer this advantage.  The electronic basis (\ref{elec-basis})
is independent of $R$ and is identical to that used by Moss and
Sadler~\cite{MosSad89}.  The vibrational basis is similar to theirs in
functional form, but we used a different nonlinear parameter $\gamma$
that allowed us to avoid certain expressions involving
hypergeometric series and thereby offered an apparent improvement in
speed.  We expect that the accuracy of our vibrational basis is at
least equal to that of Moss and Sadler.


\section{CALCULATION}
Matrix elements of the Hamiltonian over the basis set functions and
the overlap between basis set functions were set up as
four-dimensional integrals over $\lambda,\mu,\chi$, and $R$. The evaluations
reduce to integrals over $\lambda$, $\mu$, and $R$.
The eigenvalues were obtained using
the Rayleigh-Ritz method by  
solution
of the generalized eigenvalue
problem for the Hamiltonian and overlap matrices
and iteratively varying the nonlinear parameters.
Some details on the integrals and procedures are
presented in this section. 
%
\subsection{Evaluation of the integrals}
Consider the integrals over $\lambda$  and over $R$ required for evaluation
of the Hamiltonian and overlap matrix elements.
Any integrals containing derivatives were manipulated 
to eliminate the derivatives by utilizing 
\begin{equation}
\frac{\partial}{\partial x}L_n^a (x) =  -L_{n-1}^{a+1} (x)
\end{equation}
and
\begin{equation}
	\label{Lag-lower}
        L_{n}^{a} (x) = \sum_{k=0}^{n} L_{k}^{a-1} (x)
\end{equation}
to rewrite each integrand as a linear combination of integrals of the
form
\begin{equation}
\label{primary}
\int_0^\infty dx\,x^{a+r} L_m^a (x)L_n^a (x) e^{-x},
\end{equation}
where $r$ is an integer, $r\geq 0$.

The resulting sets of integrals of form~(\ref{primary}),
and any other integrals of that form, were
then manipulated to eliminate the powers of $\lambda$.
This was done by writing the product 
$x^{r} L_m^a (x)$ as a linear 
combination of
Laguerre polynomials with the same superscript.
To this end, the expression
\begin{equation}
x L_n^a (x) = (n+a) L_n^{a-1} (x) - (n+1) L_{n+1}^{a-1} (x),
\end{equation}
derived using the summation definition for associated
Laguerre polynomials,
can be reduced using
\begin{equation}
\label{Lag-raise}
 L_n^a (x)  = L_n^{a+1} (x) - L_{n-1}^{a+1} (x)
\end{equation}
to the desired expression,
\begin{equation}
\label{simple-xlag}
x L_n^a (x)
  = (2n+a+1) L_n^{a} (x) - (n+1) L_{n+1}^{a} (x) - (n+a)L_{n-1}^a (x).
\end{equation}

Substituting Eq.~(\ref{simple-xlag}) into Eq.~(\ref{primary}),
each integral over $\lambda$ can now be expressed as a sum of integrals of
the form
\begin{equation}
\label{orthog}
\int_0^\infty dx\,x^{a} L_m^a (x)L_n^a (x) e^{-x}
 = \delta_{mn} (m+a)!/m!.
\end{equation}

The integrals involving $\mu$ could be performed through simple
manipulations of associated Legendre polynomials.

Coupling between states of different $\Lambda$ introduced two
problems.  The first was 
that in order to carry out manipulations such as those used
above leading to~(\ref{orthog}), we required expressions
for raising or lowering superscripts by more than unity.
Using Eq.~(\ref{Lag-lower}) we derived the relation
\begin{equation}
	L_{n}^{a} (x) = \sum_{k=0}^{n} {{l +k-1}\choose {k} }
	L_{n-k}^{a-l} (x)
\end{equation}
and similarly from repeated application of Eq.~(\ref{Lag-raise})
we derived the relation
\begin{equation}
	L_{n}^{a} (x) = \sum_{k=0}^{l} (-1)^{k} {{l}\choose{k}}
	L_{n-k}^{a+l} (x).
\end{equation}

The second problem was the coupling of different $\gamma$
parameters. By using the same manipulations as  for the $\lambda$
integral, we  reduce the vibrational integral to a linear
combination of functions $I$, where
\begin{equation}
I (a,m,n,\gamma_i,\gamma_j)\equiv \int_0^\infty dx\, x^a
 L_m^a (\gamma_i x) L_n^a (\gamma_j x) \exp (-\case{1}{2}
  (\gamma_i+\gamma_j) x) ,
\end{equation}
which can be reexpressed in terms of the hypergeometric function ${}_2
F_1$ using Eq.~(7.414.4) of Ref.~\cite{GraRyz94} as
\begin{equation}
\label{secondary}
I (a,m,n,\gamma_i,\gamma_j)  =
F (-m,-n;-m-n-a;\gamma_{\rm rat}^2) \frac{(m+n+a)!}{m!n!} 2^{a+1}
  (-1)^m \gamma_{\rm rat}^{-n-m} (\gamma_i+\gamma_j)^{-a-1} ,
\end{equation}
where
\begin{equation}
\gamma_{\rm rat}\equiv(\gamma_i+\gamma_j)/(\gamma_i-\gamma_j).
\end{equation}
The hypergeometric series
terminates
since $m\geq 0$ and $n\geq 0$.
Some additional notes on evaluating integrals of Laguerre
and Legendre polynomials are given in~\cite{MosSad89}. 
Maple V was used to check the evaluation of the matrix elements and it
was used to output them into Fortran code.

\subsection{Numerical procedures}\label{subsec:numerical}

The trial functions~(\ref{tf}) have three sectors.
They are  comprised of two
electronic sectors, labeled by the indices $b$ and $c$ and governed by
the nonlinear parameter $\alpha$, and one vibrational sector, labeled
by the index $d$ and governed by the nonlinear parameters $\beta$ and
$\gamma$.  In our calculations  each
sector was treated separately in optimizing the nonlinear
parameters and in studying convergence as the basis size was
increased.  The eigenvalues and wave functions were determined by
solution of the secular equation using the {\sc lapack} routines DSYGV and
DSPGV, part of the math subroutine library {\sc dxml}.  The energy was
further minimized by iteratively varying various nonlinear parameters
(using a procedure discussed below)
and rediagonalizing. For small basis set sizes we used a conjugate
gradient method and then minimized by hand  and  for the larger basis set
sizes we used an algorithm similar to Brent's~\cite{NumRec}.
Minimization of $\alpha$ was accomplished with standard algorithms. The
optimum values for the parameters $\beta$ and $\gamma$ were more difficult to
determine for two reasons. First, $\beta$ is integer and the
necessarily discrete choices impeded the optimization; furthermore, a
change in $\beta$ does not correspond to a parabolic change in the
value of the energy.  Second, the nonlinear parameters $\beta$
and $\gamma$ are intrinsically linked requiring simultaneous minimization.

A general procedure was developed which allowed us to optimize
$\alpha$, $\beta$, and $\gamma$ efficiently.  Four steps can be
identified.  1)~We fixed $\beta$ and $\gamma$ and then $\alpha$ was
optimized for a minimum energy. 2)~To minimize on $\beta$ and $\gamma$
we fixed $\beta$ and then minimized on $\gamma$. The
parameter $\beta$ was then
varied by a large interval (about 6) and then we minimized again on
$\gamma$.  Some care was required in selecting what would be
the optimum values of $\gamma$ as false local minima occasionally
appeared.  3)~Values of $\beta$ within the final interval
were searched for the optimum value with minimization on $\gamma$.
4)~After all of the above $\alpha$ was reoptimized with the selected
$\beta$ and $\gamma$. In all cases it was found in step~4) that the
value of $\alpha$ was the same as that  found in step~1),
an important verification of our choice of final
optimized nonlinear parameters.

\begin{figure}[p]
\epsfxsize=1.\textwidth \epsfbox{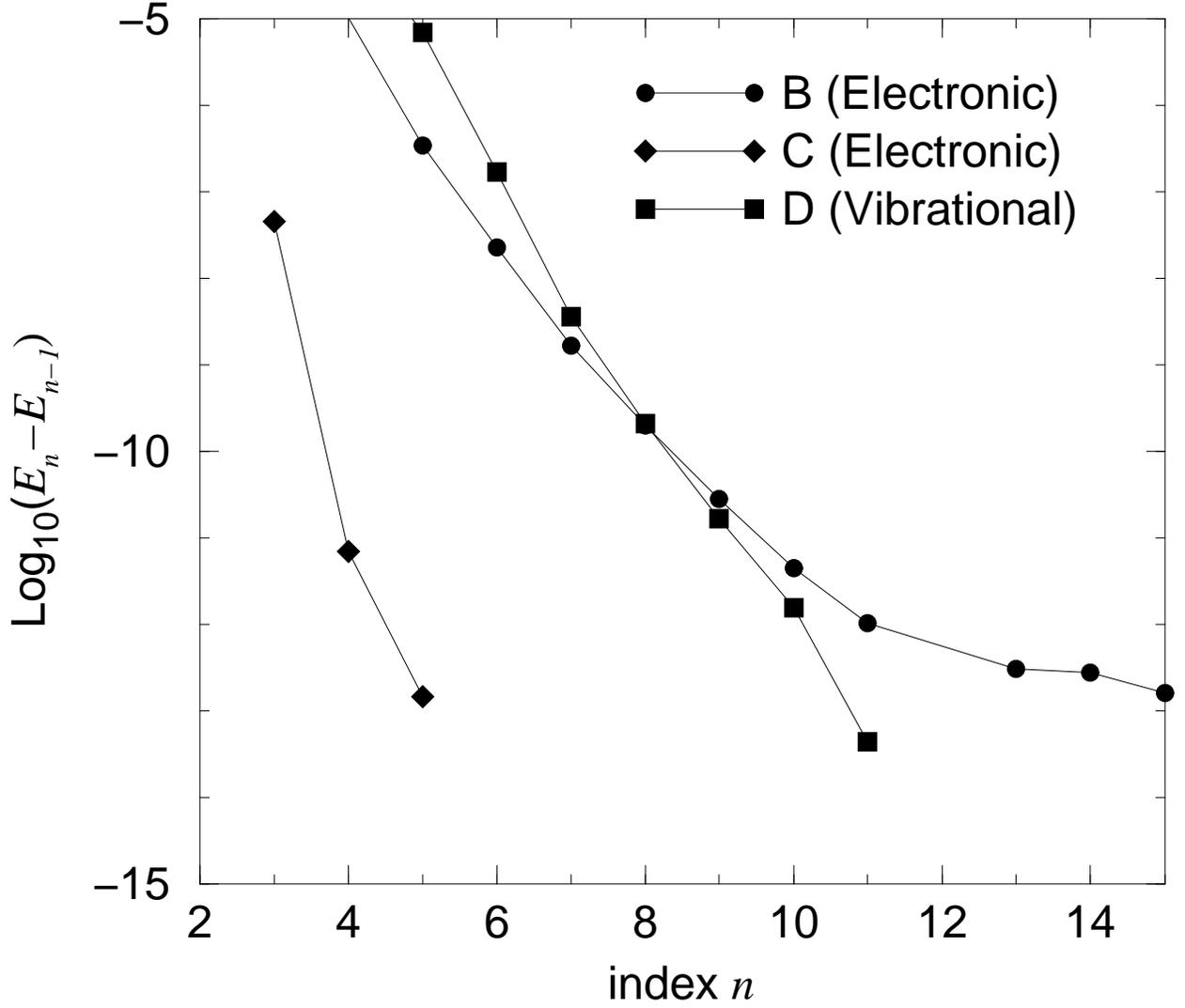}
\caption{Convergence study for the ground state $\Sigma_g$ energy of
$\mbox{H}_2{}^+$ with $v=0,N=0$.  The three basis sectors are fixed at
their optimized dimensions for $B$, $C$, and $D$.  Then for each
sector, in turn, the index of the basis set $B$, $C$, or $D$, is set
back to 2 and the value is increased until the optimized value of $B$,
$C$, or $D$ is reached again.  Each line represents the $\log_{10}$ of
the energy for the index value $n$ subtracted from the energy for the
previous index value.  (For sector $B$ we have omitted the energy
$E_{12}$.)
\label{sigma-g-fig}}
\end{figure}

\begin{figure}[p]
\epsfxsize=1.\textwidth \epsfbox{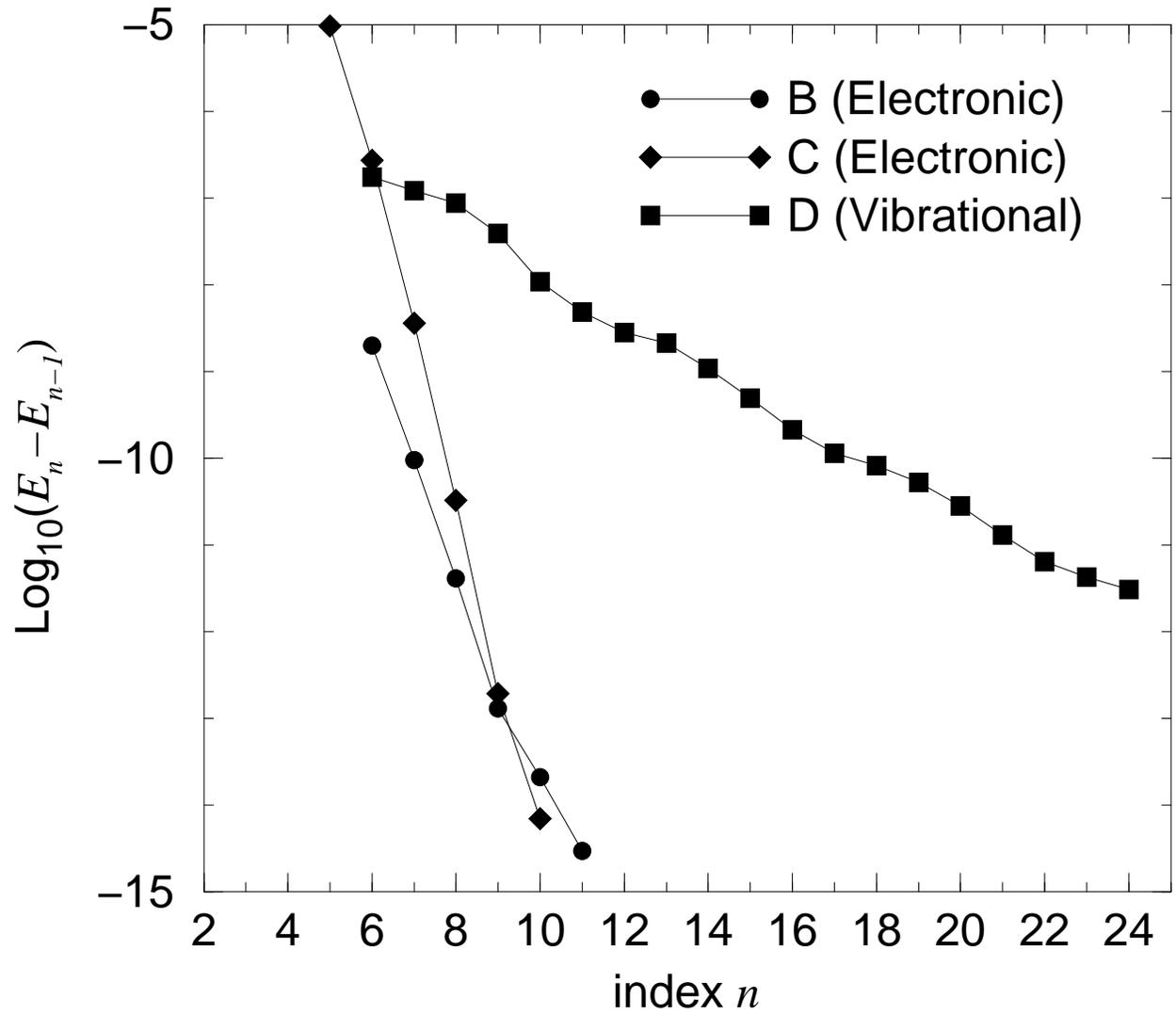}
\caption{Convergence study for the $\Sigma_u$ energy of $\mbox{H}_2{}^+$
with $v=0,N=0$.
\label{sigma-u-fig}}
\end{figure}

\begin{figure}[p]
\epsfxsize=1.\textwidth \epsfbox{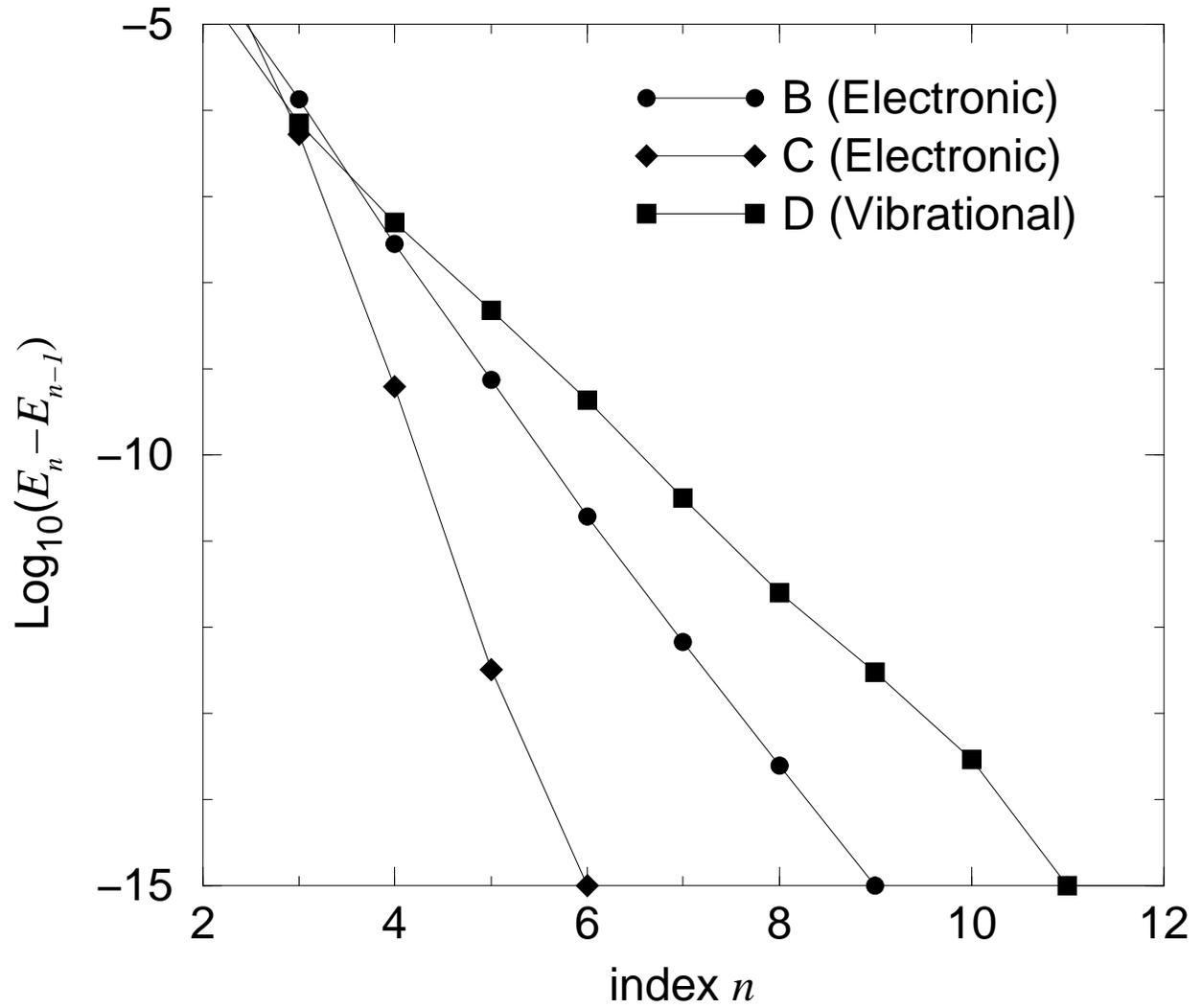}
\caption{Convergence study for the $\Pi_u$ energy of $\mbox{H}_2{}^+$ for the
$v=0, N=1$ state with with no coupling to the $\Sigma_u$ symmetry
included.
\label{pi-u-fig}}
\end{figure}

Having fixed the nonlinear parameters the basis set size was
systematically increased to obtain precise eigenvalues by expanding
each sector separately.  Convergence to the final value was
logarithmic.  For   $\mbox{H}_2{}^+$ 
in Figs.~\ref{sigma-g-fig}, \ref{sigma-u-fig}, and
\ref{pi-u-fig}  the
convergence is demonstrated by plotting the difference between the
energy for a particular basis set dimension and the energy for a basis
set of dimension one unit larger. 
Results for $\mbox{D}_2{}^+$ are similar.
For each figure, we begin with the
final optimized wave function.  The nonlinear parameters are not
changed but the basis set dimension is set to $B=2$, then index $B$
is increased with the others held fixed at their optimized values and
the difference between successive energies is plotted yielding the
curves labeled ``$B$ (Electronic)'' and similarly for $C$ and $D$.
For the $\Sigma_u$ states of $\mbox{H}_2{}^+$ and
$\mbox{D}_2{}^+$ convergence in the vibrational sector is slower than
for the $\Sigma_g$ and $\Pi_u$ states so we extrapolated to the
desired numerical accuracy using linear regression on the log of the
energy differences.  Figure~\ref{sigma-u-fig} 
illustrates the slow convergence but also the validity of the
extrapolation.  The basis set dimensions and nonlinear parameters for
states with $N=0$ are given in Table~\ref{opt-table} for $\Sigma_g$
symmetry in the first row under ``Type I'' and for $\Sigma_u$ symmetry
in the first row under ``Type II''.

\begin{figure}[p]
\epsfxsize=1.\textwidth \epsfbox{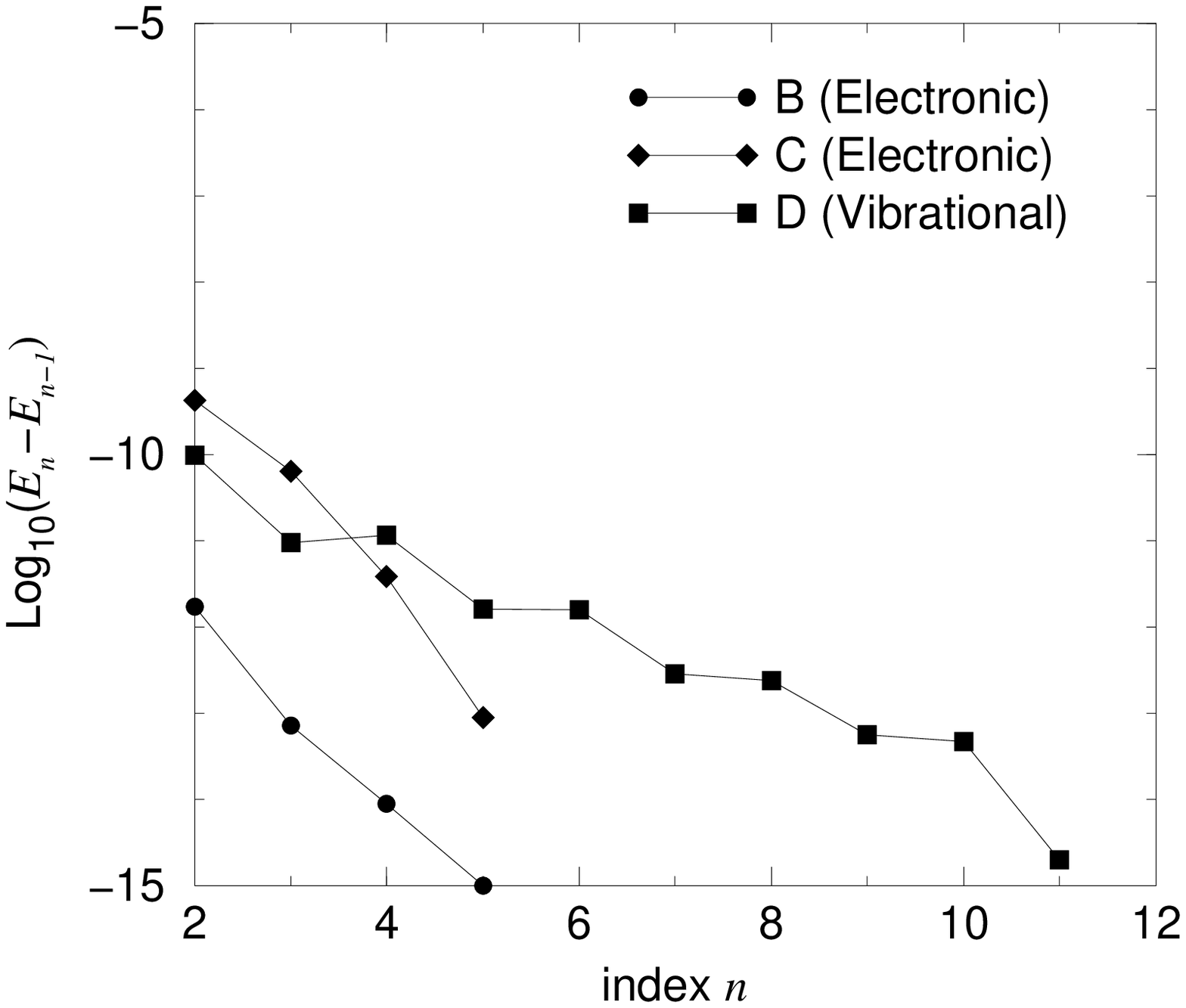}
\caption{Convergence study for the energy
of $\mbox{H}_2{}^+$ in the $\Sigma_u$, 
$v=0, N=1$ state 
for the basis set of $\Pi_u$  symmetry
entering in the calculation.
The $\Sigma_u$ symmetry basis set is fixed with the optimized
size and nonlinear parameters listed
in Table~\protect\ref{opt-table} for the calculations
of this plot.
\label{coupled-sigma-u-fig}}
\end{figure}

\begin{figure}[p]
\epsfxsize=1.\textwidth \epsfbox{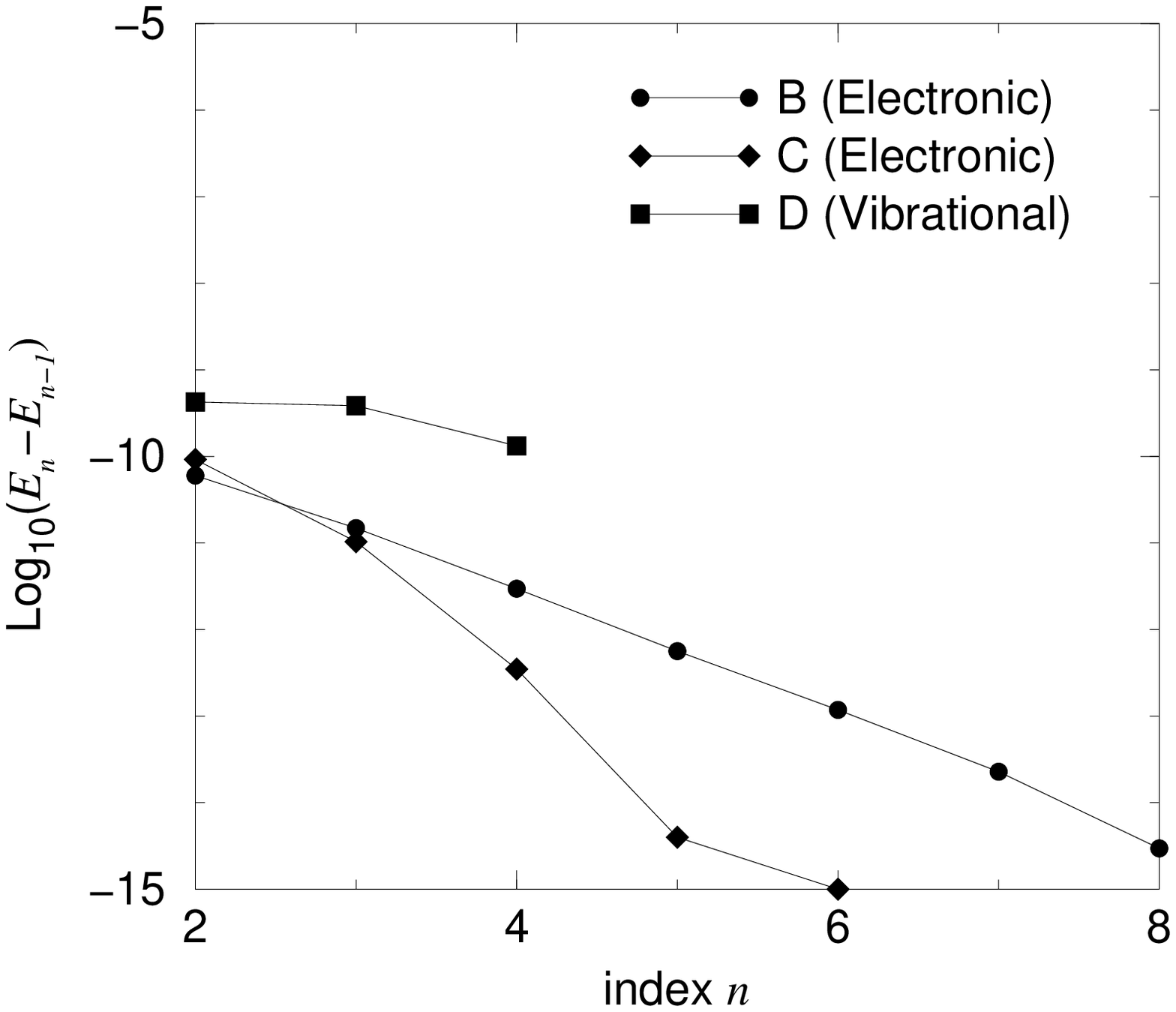}
\caption{
Convergence study for the energy
of $\mbox{H}_2{}^+$ in the $\Pi_u$, 
$v=0, N=1$ state 
for the basis set of $\Sigma_u$  symmetry
entering in the calculation.
The $\Pi_u$ symmetry basis set is fixed with the optimized
size and nonlinear parameters listed
in Table~\protect\ref{opt-table} for the calculations
of this plot.
\label{coupled-pi-u-fig}}
\end{figure}

For the states with $N>0$, the off-diagonal term Eq.~(\ref{sigpi}) in
the Hamiltonian requires the inclusion of coupling between basis sets
of $\Sigma$ and $\Pi$ symmetry.  Denoting the electronic basis sets by
their value of $\Lambda$ as $|\Lambda\rangle$ we set up matrix
elements of the Hamiltonian  using the rotated basis
$\case{1}{\sqrt{2}}(|\mbox{+}1\rangle +|\mbox{$-$}1\rangle)$ and
$\case{1}{\sqrt{2}}(|\mbox{+}1\rangle -|\mbox{$-$}1\rangle)$.  With it
there is only coupling between $|0\rangle$ and
$\case{1}{\sqrt{2}}(|\mbox{+}1\rangle - |\mbox{$-$}1\rangle)$.  A two
by two matrix of matrices was created with the uncoupled Hamiltonian
matrix elements for each basis set as the diagonal elements and the
matrix elements of the coupling term Eq.~(\ref{sigpi}) between the two
basis sets as the off-diagonal elements.  The energies of the states
were determined by diagonalization of this matrix, while the energies
corresponding to the uncoupled basis
$\case{1}{\sqrt{2}}(|\mbox{+}1\rangle +|\mbox{$-$}1\rangle)$ were
determined by diagonalization of the uncoupled Hamiltonian.  For each
state, the non-linear parameters and basis size were fixed at the
values already determined for the minimum energies.  Then the same
technique used for the uncoupled energies was applied to the coupled
basis sets to determine non-linear parameters and basis sizes that
minimized the energy of the state under consideration.  For example,
when trying to determine the $\Sigma_u,v=0,N=1$ energy, the $\Sigma$
basis set parameters were held fixed at their uncoupled values, and
the $\Pi$ basis set parameters were changed.  The parameters for the
coupling basis set were significantly different from those which
minimized the energy in the uncoupled calculations, requiring six
specialized parameters for each state when coupling was considered.
The rate of convergence of the coupling terms is illustrated in
Figs.~\ref{coupled-sigma-u-fig} and~\ref{coupled-pi-u-fig} 
for $\mbox{H}_2{}^+$. The
energies converge logarithmically as each sector dimension is
increased in turn. To evaluate the contribution of this small
off-diagonal term to the energy many fewer basis set elements are
needed than for the diagonal terms.  The basis set dimensions and
nonlinear parameters for states with $N>0$ are given in
Table~\ref{opt-table}.  For each symmetry there are two rows. The
first row lists the dimensions and parameters for the primary symmetry
used for all calculations 
and the second row lists the quantities for the 
additional symmetry required for $N>0$ entering
through the coupling of Eq.~(\ref{sigpi}).

The total number of basis functions used can be calculated from the
data listed in Table~\ref{opt-table} and is
the sum of the values of $S$ defined in Eq.~(\ref{tf}) entering for
each symmetry. For example, for
$\mbox{H}_2{}^+$ $\Sigma_g$, $v=0$, $N=0$, we used
$(13+1)\times(5+1)\times (13+1)=1176$ functions and
for $N=1$ we used $1176 + (5+1)\times (4+1)\times (6+1)=1386$
functions.
For
$\mbox{H}_2{}^+$  $\Pi_u$, $v=0$, $N=1$ we used two runs, 
each corresponding to one 
of the rotated basis sets.  For the uncoupled set, we had $910$ functions,
while for the coupled set, we used $910 + 270 = 1180$ functions.


\section{DISCUSSION}

Tables~\ref{e-h-sig-table} and~\ref{e-d-sig-table} compare the present
calculations of nonadiabatic energies for $\mbox{H}_2{}^+$ and
$\mbox{D}_2{}^+$ respectively with available precision calculations.
In each table the vibration-rotation eigenvalues for the $\Sigma_g$
symmetry are given first, followed by those for the $\Sigma_u$
symmetry.

For the  $\Sigma_g$ state 
the most precise variational basis set calculations are given
for $\mbox{H}_2{}^+$
in Refs.~\cite{BisChe77b,BisSol85,GreDelBil98,Mos93b} 
and for $\mbox{D}_2{}^+$
in Refs.~\cite{BisChe77b,BisSol85,Mos93b}.
Variation-perturbation calculations have been performed
by
Wolniewicz and Orlikowski~\cite{WolOrl91} for 
$\mbox{H}_2{}^+$ and $\mbox{D}_2{}^+$ for all the $\Sigma_g$
vibration-rotation 
states but the tabulated results include radiative
and relativistic corrections and can not be compared directly with the
present work.  Using the artificial channel approach Moss carried out
extensive nonadiabatic calculations of all the vibrational-rotational
states of $\mbox{H}_2{}^+$~\cite{Mos93b} and
$\mbox{D}_2{}^+$~\cite{Mos93a} for the $\Sigma_g$  states.  His
results with radiative and relativistic corrections are in good
agreement with Wolniewicz and Orlikowski and he also presented
energies without these corrections. 
In 
Tables~\ref{e-h-sig-table} and~\ref{e-d-sig-table}
the various calculations for the $v=0,N=0$, $v=0,N=1$, and $v=1,N=0$ 
states are compared to our calculations.
Results listed in
Refs.~\cite{Mos93a,Mos93b} are converted from dissociation
energies in wavenumbers to atomic units and combined with the
asymptotic energy $-M_n/[2 (1+M_n)]$.  
Our results are consistent with and slightly
improve upon the precision of previous calculations.

Only a few high-precision calculations are available for the lowest
states of $\Sigma_u$ symmetry for $\mbox{H}_2{}^+$ and
$\mbox{D}_2{}^+$.  Wolniewicz and Orlikowski used the
variation-perturbation method and found 3 bound levels for
$\mbox{H}_2{}^+$ and 7 bound levels for $\mbox{D}_2{}^+$ and gave
energies of the levels with $\Sigma$-$\Pi$ coupling included.
Subsequently, Moss using the artificial channel method including
$\Sigma$-$\Pi$ coupling found results in agreement with those of
Wolniewicz and Orlikowski for both $\mbox{H}_2{}^+$~\cite{Mos93b} and
$\mbox{D}_2{}^+$~\cite{Mos93a}.  Our $\Sigma_u$ results are compared
with these prior calculations in Tables~\ref{e-h-sig-table}
and~\ref{e-d-sig-table}.  For the $v=0,N=0$ 
and $v=0,N=1$ states our energies are consistent with the others
and  of
higher precision.
However, for the $\mbox{D}_2{}^+$ $v=1,N=0$ state we found that
a quite large basis set ($B=20, C=11, D=36$ with $\alpha=15.8$,
$\beta=37$ and $\gamma=2.6$) was required to approach the energies
given in Refs.~\cite{WolOrl91,Mos93a}.
Peek~\cite{Pee69} showed that in the Born-Oppenheimer approximation
the $v=1,N=0$ vibrational wave function can have significant
amplitude at values of $R$ as large as several hundred $a_0$.
Our electronic basis set is not explicitly dependent on $R$
and this may account for the large basis size needed.
Other methods~\cite{WolPol86,WolOrl91,Mos93a,Mos93b} are based
on coupled channel approaches that may be better
at describing such diffuse vibrational states.

There do not appear to be any published nonadiabatic energies for the
lowest electronic state of $\Pi_u$ symmetry of either $\mbox{H}_2{}^+$ or
$\mbox{D}_2{}^+$.  Probably the most accurate study published is that of Bishop
{\em et al.\/}~\cite{BisShiBec75}, who investigated the $\Pi_u$
energies of $\mbox{H}_2{}^+$ within the standard adiabatic
approximation~\cite{Kol69,BisWet73}.  In Table~\ref{pi-table} the
present nonadiabatic energies are compared to Born-Oppenheimer and
standard adiabatic energies.  The energy calculated in the
Born-Oppenheimer approximation is a lower bound to the true energy
while the standard adiabatic and nonadiabatic energies are upper
bounds~\cite{Eps66,HunGraPri66}.  The standard
adiabatic energies were calculated with the diagonal coupling of
Ref.~\cite{BisShiBec75} rescaled to a proton mass of $1\,836.152\,701$
and the results differ in the seventh decimal place from the values
reported in~\cite{BisShiBec75}. The present nonadiabatic results 
lie above the Born-Oppenheimer energy but below the standard
adiabatic energy as expected~\cite{HunGraPri66}.

The energies in Table~\ref{pi-table} were calculated without
the consideration of Eq.~(\ref{sigpi}) leading
to one level for each value of $N$.
With the inclusion of the coupling term (\ref{sigpi}) as
described above in Sec.~\ref{subsec:numerical}
our calculations exhibit lambda-doubling in the eigenvalues of $\Pi$
symmetry.  
In Table~\ref{e-h-pi-table} 
calculated eigenvalues for the $v=0$ and 1 states with $N=1$
are presented for $\mbox{H}_2{}^+$ and $\mbox{D}_2{}^+$.
For each value of $v$ the first row gives the energy
of the shifted level resulting from the diagonalization of
the matrix coupling  $|0\rangle$ and 
$\case{1}{\sqrt{2}}(|\mbox{+}1\rangle - |\mbox{$-$}1\rangle)$
and the second row gives the energy of the other, unshifted, level.
The energy difference between the two levels is the lambda-doubling.

%
\acknowledgements
We are grateful to Prof. P. Froelich, Dr. S. Jonsell,
and Prof. J. Shertzer for helpful comments.
This work was supported in part by the U.S. Department of Energy,
Division of Chemical Sciences, Office of Basic Energy Sciences, Office
of Energy Research.  
ZCY was also supported by the Natural Sciences and Engineering
Research Council of Canada.
The Institute for Theoretical Atomic and
Molecular Physics is supported by a grant from the National Science
Foundation to the Smithsonian Institution and Harvard University.
%
\begin{table}
\begin{center}
\caption{For $\mbox{H}_2{}^+$ values of the dimensions $B$, $C$, and $D$
and the optimized nonlinear parameters $\alpha$, $\beta$, and $\gamma$.
The values used for $\mbox{D}_2{}^+$ are identical except for 
the three values listed
in parentheses.}
\label{opt-table}
\begin{tabular}{cccccccc}
\multicolumn{2}{c}{} &
   \multicolumn{3}{c}{Dimension}&
   \multicolumn{3}{c}{Nonlinear parameter}  \\
\multicolumn{1}{c}{Type} &
   \multicolumn{1}{c}{Symmetry} &
   \multicolumn{1}{c}{$B$} &
   \multicolumn{1}{c}{$C$} &
   \multicolumn{1}{c}{$D$} &
   \multicolumn{1}{c}{$\alpha$} &
   \multicolumn{1}{c}{$\beta$} &
   \multicolumn{1}{c}{$\gamma$} \\
\hline
I & $\Sigma_g$ &  13 &  5 &  13(17) &  3.1561 & 67  &  37.0  \\
  & $\Pi_g$    &   5 &  4 &   6     &  3.0    & 79  &  42.0  \\
\multicolumn{8}{c}{}   \\		        	  
II & $\Sigma_u$&14(10) &  11(9) &  30     &  15.8   & 43  &   3.1  \\
   & $\Pi_u$   &   5   &  5     &  11     &  13.0   & 97  &   7.4  \\
\multicolumn{8}{c}{}   \\		        	  
III& $\Pi_u$   &   9 &  6 &  12(19) &  6.0    & 125 &  16.5  \\
   & $\Sigma_u$&   8 &  5 &  4      &  5.0    & 47  &  3.86  \\
\end{tabular}
\end{center}
\end{table}
%
%
%
\begin{table}
\begin{center}
\caption{ Comparison of nonadiabatic vibration-rotation energies for
$\mbox{H}_2{}^+$ for each of the lowest electronic states of $\Sigma_g$ or
$\Sigma_u$ symmetry.  Calculations 
with $N>0$ include the coupling term of
Eq.~(\protect\ref{sigpi}).  Unless indicated otherwise all
calculations correspond to a proton mass of $1\,836.152\,701$ in units of
the electron mass.}
\label{e-h-sig-table}
\begin{tabular}{llll}
\multicolumn{1}{c}{State} &\multicolumn{1}{c}{Author (Year)} &
   \multicolumn{1}{c}{Ref.}  & \multicolumn{1}{c}{Energy} \\
\hline
$\Sigma_g, v=0,N=0$  
     & Bishop and Cheung (1977)\tablenote{Proton mass 1836.15}
                & \cite{BisChe77b} & $-$0.597\,139\,062\,5 \\
     & Bishop and Solunac (1985)\tablenotemark[1]{}
                & \cite{BisSol85} & $-$0.597\,139\,063\,18 \\
     & Moss (1993) & \cite{Mos93b}  & $-$0.597\,139\,063\,1 \\
     & Gr{\'e}maud et al. (1998) & \cite{GreDelBil98} &
                                        $-$0.597\,139\,063\,123(1) \\
     & This work &  & $-$0.597\,139\,063\,123\,9(5) \\
$\Sigma_g, v=0,N=1$  
     & Moss (1993) & \cite{Mos93b}  & $-$0.596\,873\,738\,9 \\
     & This work &                  & $-$0.596\,873\,738\,832\,8(5) \\
$\Sigma_g, v=1,N=0$  
     & Bishop and Cheung (1977)\tablenotemark[1]{}
            & \cite{BisChe77b} & $-$0.587\,155\,675\,8 \\
     & Moss (1993)     & \cite{Mos93b}             & $-$0.587\,155\,679\,2 \\
     & Gr{\'e}maud et al. (1998) & \cite{GreDelBil98} &
                      $-$0.587\,155\,679\,212(1) \\
     & This work &  & $-$0.587\,155\,679\,213\,6(5) \\
\multicolumn{4}{c}{ }\\
$\Sigma_u, v=0,N=0$   
     & Wolniewicz and Orlikowski (1991)&\cite{WolOrl91} & $-$0.499\,743\,49\\
     & Moss (1993) & \cite{Mos93b} & $-$0.499\,743\,502\,2 \\
     & This work &                 & $-$0.499\,743\,502\,21(1) \\
$\Sigma_u, v=0,N=1$   
     & Wolniewicz and Orlikowski (1991)&\cite{WolOrl91} &$-$0.499\,739\,25\\
     & Moss (1993) & \cite{Mos93b} & $-$0.499\,739\,268\,0 \\
     & This work &                 & $-$0.499\,739\,267\,93(2)\tablenote{For 
this energy, the basis set had dimension $B=16$.}
\end{tabular}
\end{center}
\end{table}
%
%
\begin{table}
\begin{center}
\caption{ Comparison of nonadiabatic vibration-rotation energies for
$\mbox{D}_2{}^+$ for each of the lowest electronic states of $\Sigma_g$
or $\Sigma_u$ symmetry.  Calculations 
with $N>0$ include the coupling term of
Eq.~(\protect\ref{sigpi}).  Unless indicated otherwise all
calculations correspond to a deuteron mass of $3\,670.483\,014$ in units of
the electron mass.}
\label{e-d-sig-table}
\begin{tabular}{llll}
\multicolumn{1}{c}{State} &\multicolumn{1}{c}{Author (Year)} &
   \multicolumn{1}{c}{Ref.}  & \multicolumn{1}{c}{Energy} \\
\hline
$\Sigma_g, v=0,N=0$  
      & Bishop and Cheung (1977)\tablenote{Deuteron mass 3670.48} 
              & \cite{BisChe77b} & $-$0.598\,788\,782\,0 \\
     & Bishop and Solunac (1985)\tablenotemark[1]{}
              & \cite{BisSol85} & $-$0.598\,788\,782\,22 \\
      & Moss (1993) & \cite{Mos93a}  & $-$0.598\,788\,784 \\
      & This work &  & $-$0.598\,788\,784\,330\,8(1) \\
$\Sigma_g, v=0,N=1$  
      & Moss (1993) & \cite{Mos93b}  & $-$0.598\,654\,873\,1 \\
      & This work &                  & $-$0.598\,654\,873\,220\,5(5) \\
$\Sigma_g, v=1,N=0$  
      & Bishop and Cheung (1977)\tablenotemark[1]{}
              & \cite{BisChe77b} & $-$0.591\,603\,115\,4 \\
      & Moss (1993)     & \cite{Mos93a}  & $-$0.591\,603\,122 \\
      & This work &                      & $-$0.591\,603\,121\,903\,2(1) \\
\multicolumn{4}{c}{ }\\
$\Sigma_u, v=0,N=0$   
      & Wolniewicz and Orlikowski (1991)&\cite{WolOrl91} & $-$0.499\,888\,93\\
      & Moss (1993) & \cite{Mos93a} & $-$0.499\,888\,937\,5 \\
      & This work &                 & $-$0.499\,888\,937\,71(1) \\
$\Sigma_u, v=0,N=1$   
      & Wolniewicz and Orlikowski (1991)&\cite{WolOrl91} & $-$0.499\,886\,38 \\
      & Moss (1993) & \cite{Mos93a} & $-$0.499\,886\,382\,5 \\
      & This work &                 & $-$0.499\,886\,382\,63(1) \\
$\Sigma_u, v=1,N=0$   
      & Wolniewicz and Orlikowski (1991)&\cite{WolOrl91} & $-$0.499\,865\,21 \\
      & Moss (1993) & \cite{Mos93a} & $-$0.499\,865\,221\,0 \\
      & This work   &  & $-$0.499\,865\,217\,(5)\tablenote{For this energy, the basis set had dimensions $B=20$, $C=11$, $D=36$ with nonlinear parameters $\alpha=15.8$, $\beta=37$, and $\gamma=2.6$ as discussed in the text.}  
\end{tabular}
\end{center}
\end{table}

\clearpage
%
%
\begin{table}
\begin{center}
\caption{For $\mbox{H}_2{}^+$ the first several eigenvalues of the $\Pi_u$
symmetry with $N=1$ calculated nonadiabatically compared with
Born-Oppenheimer and standard adiabatic calculations, respectively.
For the present calculations, col. 4, the coupling term~(\ref{sigpi}) has
not been included.}
\label{pi-table}
\begin{tabular}{cccc}
\multicolumn{1}{c}{Vibrational state}   &
   \multicolumn{1}{c}{Born Oppenheimer} &
   \multicolumn{1}{c}{Standard Adiabatic} &
   \multicolumn{1}{c}{Present\protect\tablenote{Nonlinear 
   parameters $\alpha=6.0,\beta =125,\gamma=16.5$
   with $B=9, C=6, D=24$. 
}}  \\
\hline
0  & $-$0.133\,905\,216\,5 & $-$0.133\,841\,244\,8 & $-$0.133\,841\,939\,2 \\  
1  & $-$0.132\,752\,851\,6 & $-$0.132\,689\,153\,4 & $-$0.132\,689\,769\,1 \\
2  & $-$0.131\,660\,981\,7 & $-$0.131\,597\,475\,8 & $-$0.131\,598\,133\,6 \\
3  & $-$0.130\,631\,351\,9 & $-$0.130\,567\,953\,2 & $-$0.130\,568\,676\,9 \\
4  & $-$0.129\,666\,127\,2 & $-$0.129\,602\,748\,3 & $-$0.129\,603\,541\,6 \\
\end{tabular}
\end{center}
\end{table}
%
%
%
\begin{table}
\begin{center}
\caption{Lambda-doubling in nonadiabatic vibration-rotation energies
of $\mbox{H}_2{}^+$ 
and $\mbox{D}_2{}^+$ 
for the lowest electronic state of $\Pi_u$
symmetry for $v=0$ and 1, with $N=1$.  For each value of $v$ the first
row gives the energy of the shifted level arising from the coupling
term in Eq.~(\protect\ref{sigpi}) and the second row gives the energy
of the other, unshifted, level.}
\label{e-h-pi-table}
\begin{tabular}{lll}
\multicolumn{1}{c}{Ion} &
\multicolumn{1}{c}{State} &  \multicolumn{1}{c}{Energy} \\
\hline
$\mbox{H}_2{}^+$ &$\Pi_u,v=0,N=1$  &  $-$0.133\,841\,940\,395(5) \\
                 &                 &  $-$0.133\,841\,939\,176\,3(1) \\
                 &$\Pi_u,v=1,N=1$  &  $-$0.132\,689\,769\,820(5) \\
                 &                 &  $-$0.132\,689\,769\,121\,8(1) \\
$\mbox{D}_2{}^+$ &$\Pi_u,v=0,N=1$  & $-$0.134\,052\,118\,044(5) \\
                 &                 & $-$0.134\,052\,117\,739\,8(1) \\
                 &$\Pi_u,v=1,N=1$  & $-$0.133\,224\,515\,520(5) \\
                 &                 & $-$0.133\,224\,515\,448\,7(1) \\
\end{tabular}
\end{center}
\end{table}
\clearpage
\clearpage
\clearpage
\clearpage
%
%

%
%
\end{document}